# Metal-Ferroelectric-Metal heterostructures with Schottky contacts
# I. Influence of the ferroelectric properties

L. Pintilie[a]

NIMP, P.O. Box MG-7, 077125 Bucharest-Magurele, Romania
and
Max Planck Institute of Microstructure Physics, Weinberg 2, 06120 Halle, Germany

M. Alexe
Max Planck Institute of Microstructure Physics, Weinberg 2, 06120 Halle, Germany

## *Abstract*
A model for Metal-Ferroelectric-Metal structures with Schottky contacts is proposed. The model adapts the general theories of metal-semiconductor rectifying contacts for the particular case of metal-ferroelectric contact by introducing: the ferroelectric polarization as a sheet of surface charge located at a finite distance from the electrode interface; a deep trapping level of high concentration; the static and dynamic values of the dielectric constant. Consequences of the proposed model on relevant quantities of the Schottky contact such as built-in voltage, charge density and depletion width, as well as on the interpretation of the current-voltage and capacitance-voltage characteristics are discussed in detail.

## *Introduction*
In the last years ferroelectric thin films, especially with perovskite structure, have attracted much interest due to their application in various microelectronic devices such as non-volatile memory cells.[1,2] For most of these applications an important factor to be addressed is the leakage current due to the presumed link between the free carriers and some detrimental phenomena in ferroelectrics such as fatigue or imprint.

Several models and transport mechanisms were proposed to explain the current-voltage characteristics of metal-ferroelectric-metal (MFM) capacitors. Screening the vast literature on this topic [3-20] the transport mechanisms can be grouped in two major categories: i) interface controlled mechanisms based on Schottky emission or Fowler-Nordheim tunneling and ii) bulk controlled mechanisms such as ohmic, space charge limited currents, Pool-Frenkel emission, ionic conduction, or a combination of them. However, the main dilemma arises from the question whatever the ferroelectric film should be treated as an insulator or as a semiconductor. This aspect is very important in calculation of the effective electric field in case of interface controlled conduction mechanisms.[21,22]

Traditionally, ferroelectrics have been regarded as insulators. This approach has been successful in explaining the basic ferroelectric properties such as phase transition and ferroelectric switching, but fails totally in explaining the transport mechanisms or other properties that involve internal mobile charges. On the other side, considering ferroelectrics as simple semiconductors have brought a new set of problems into an already complicated picture. For instance the non-doped PZT films, i.e. as close as possible to the stoichiometric composition, should be p-type according to the defect chemistry, but often they are regarded as an n-type semiconductor due to a higher mobility of the electrons.[1,21,22] Moreover, a variation of the semiconductor type within the film thickness has also been considered. Due to oxygen vacancy accumulation an n-type layer is supposed to occur at the surface whereas the bulk remains p-type.[1,23] The estimated values of the charge concentration are spread over a wide range, i.e. from $10^{17}$ cm$^{-3}$ to $10^{21}$ cm$^{-3}$, depending on the calculation method and the physical model used.[1,24-27] It is worth to mention that the origin and properties of the so-called "dead layer", which is often used in the serial capacitor model to explain the thickness dependence of the dielectric permittivity, is also unclear.[28-32] Therefore, it will be essential to define a sound model of the metal-ferroelectric interface that will be able to properly describe the current-voltage (I-V) and capacitance-voltage (C-V) characteristics within the same formalism.

---

[a] E-mail address: pintilie@mpi-halle.mpg.de

The present paper proposes a model for the metal-ferroelectric interfaces. The model is based on the already known theories developed for the classical case of metal-semiconductor Schottky contacts. These well-known theories are here adapted to the particular case of metal-ferroelectric interfaces considering the ferroelectric material as a wide-band gap semiconductor that additionally exhibits a ferroelectric polarization. The current-voltage (I-V) and capacitance-voltage (C-V) characteristics are then discussed in the framework of the model.

## *Metal-Ferroelectric Interface*

The present model has been suggested by the observation that the current measurements are more sensitive to the space charge near the electrodes, whereas the C-V measurements are more sensitive to the charge density at the space charge region (SCR) edge.[33,34] The model is based on the text book theories developed for metal-semiconductor rectifying contacts[35,36] and it gives special emphasis on Pb(Zr,Ti)O$_3$ (PZT) ferroelectric films, but it can be easily applied for all metal-ferroelectric systems that are similar. The following assumptions are made:

1. The ferroelectric PZT is a p-type wide-gap semiconductor. The reported values for the PZT band gap are in the 3.2-3.9 eV range, depending on the Zr/Ti ratio. The p-type character is a consequence of the defect chemistry.
2. The metal-ferroelectric contact is a rectifying (Schottky) contact. A MFM structure can be regarded as two back-to-back Schottky contacts or diodes. One of the diodes will be reversed biased irrespective of the polarity of the applied voltage. It is assumed that the ferroelectric film is thick enough to avoid the overlapping of the two depletion regions associated with Schottky contacts. There will always be in the volume of the film a region exhibiting flat bands. The two interfaces can be analyzed separately in this case, by considering the neutral volume like a virtual ohmic contact (the semi-infinite plane approximation).[35]
3. The ferroelectric polarization is modeled as an infinite sheet of surface charge located in the ferroelectric layer at a finite distance from the physical metal-ferroelectric interface. Within this thin interface layer the polarization is assumed to be zero (see Fig. 1). A MFM structure will have two sheets of surface charge, one positive and one negative, corresponding to the two faces of a poled ferroelectric. In between the two sheets of charge the polarization, considered as a vectorial quantity, is assumed uniform. Thus, no net polarization charges will be present in the ferroelectric volume.

   The interfacial layer in which the polarization vanishes can be associated with the "dead layer" proposed in various models to explain the thickness dependence of the capacitance. This is a very simple way to consider the ferroelectric polarization and its decay at the surface, compared with other models proposed for the observed size effects in ferroelectric thin films.[37]
4. The polarization hysteresis loop is assumed to be rectangular (see Fig. 2). This is the case of an ideal ferroelectric material, with an ideal switching behavior. Except at the coercive field, the ferroelectric polarization is field independent. This implies that the divergence of the ferroelectric polarization is zero even in the depleted regions where the electric field is position dependent. The above situation occurs when the polarization is set and no reversal takes place during the measurement, as is the case for the true I-V characteristic of the leakage current. The above assumption regarding the hysteresis loop has also consequences on the dielectric constant. The standard relation between the electric displacement *D*, electric field *E* and the spontaneous (or ferroelectric) polarization $P_S$ is:[38]

$$D = \varepsilon_0 \varepsilon E + P_S \qquad (1)$$

where $\varepsilon$ is the field independent dielectric constant, including the linear response of the film, and $P_S$ is the dipolar (ferroelectric) polarization. For simplicity, the ferroelectric polarization $P_S$ will be further on simply referred as "polarization" and will be noted as *P*.

From (1) it can be seen that the total dielectric constant of the ferroelectric defined as *dD/dE* includes the linear part $\varepsilon$ and the non-linear part *dP/dE*. In the case of an ideal hysteresis loop *dP/dE* =0 for any field except the coercive value. Therefore, the dielectric constant of the ferroelectric can be considered field independent and equal with the linear part $\varepsilon$.

5. The metal is considered ideal. No screening effects such as Thomas-Fermi screening are considered.
6. A deep level is assumed to exist, besides the shallow levels that are giving the conduction type (see Fig. 3). The deep level is introduced to take into account the structural defects that are usually present in ferroelectrics, and which are assumed to be uniformly distributed within the film. An acceptor-type deep level in a p-type material will contribute with charges of the same sign as the shallow acceptors, when filled in the SCR. If the deep level is donor-type, then it brings a positive contribution in the SCR and can induce an inversion of the conduction type if has a higher concentration than the shallow acceptors. For high concentrations of the deep donors, an abrupt $n^+$ - p junction can occur, which can be treated similar to a Schottky contact.. In the first approximation the type of this level, i.e. donor or acceptor, is not of major importance. What it is very important is that this level can bring a certain charge contribution in the space charge region (SCR).[34] The net charge density in the SCR is very important for the calculation of the electric field at the interface in the framework of the Schottky contact model, as will be shown later. A the deep level in high concentration can pin the Fermi level, as was assumed for the as-grown semi-insulating GaAs.[39-41] In the following analysis the deep level will be considered as acceptor-type. This assumption was suggested by a recent report on the presence of negative charges at the surface of ferroelectric films.[42]
7. Contrary to the standard assumption that the polarization charges (or the associated depolarization field) are compensated with free charges from the electrodes, in the present case it is assumed that the surface bound charges associated with the ferroelectric polarization are compensated with trapped charges and with ionized shallow impurities located in the SCR.

As can be seen from the above assumptions, additional to the conventional Schottky treatment is the introduction of ferroelectric polarization and of a deep level in the band gap of the ferroelectric layer. In the following we will investigate the above modified Schottky contact model by calculating the characteristic quantities such as built-in voltage $V_{bi}$, depletion layer width *w*, maximum electric field at the interface $E_m$, and depletion layer capacitance *C*. These characteristic quantities will later on be used to interpret the I-V and C-V characteristics.

A very similar Schottky contact model was previously proposed for a metal-semiconductor structure formed by an indium electrode on a p-type CdTe epitaxial layer, deposited on a p-type CdTe substrate of higher doping concentration.[43,44] The present model adapts this previous model to the case of metal-ferroelectric interfaces. The first step in computing the above mentioned quantities is to solve the Poisson equation for the structure presented in Fig. 1:

$$\frac{dE(x)}{dx} = \frac{\rho(x)}{\varepsilon_0 \varepsilon_{st}} \qquad (2)$$

where *E(x)* is the electric field, *ρ(x)* is the charge density, $\varepsilon_0$ is the permittivity of free space, and $\varepsilon_{st}$ is the low frequency (or static) dielectric constant. The boundary conditions for the electric field are the usual ones for a metal-semiconductor Schottky contact: *E(0) = $E_m$* (at the interface the electric field is maximum); $E(w) = 0$ (the electric field in the neutral volume is zero). Knowing the relation between the electric field and potential $E(x) = -dV(x)/dx$, the potential on the structure can be obtained by integrating the electric field over the film thickness:[43]

$$V + V_{bi} = -\frac{1}{\varepsilon_0 \varepsilon_{st}} \int_0^d \int_0^d \rho(x) dx \qquad (3)$$

where $V_{bi}$ is the built-in potential in the absence of the polarization charges, and $d$ is the film thickness. Considering that in the neutral volume the net charge is zero, $\rho(x) = 0$, the integration in (3) will extend only over the depletion layer width $w$. Assuming an appropriate Dirac function for the polarization charge and an uniform charge density in the depleted region, the integration of (3) will give the following relation that can be used to calculate the depletion width $w$:

$$V + V_{bi} = \frac{q}{2\varepsilon_0 \varepsilon_{st}} \left[ N_{eff} w^2 \mp 2\frac{P}{q}\delta \right] \qquad (4)$$

where $N_{eff}$ ($N_{eff}=N_A+N_T$) is the effective charge density in the depleted region including both contribution of the shallow acceptors $N_A$ and the deep acceptor-like levels $N_T$, $P$ is the ferroelectric polarization as a surface charge, $q$ is the electron charge, and $\delta$ is the distance between the polarization surface charge and the physical interface with the electrode (see Fig. 1). The upper sign is for the positive polarization charge, and the lower sign is for negative polarization. The significant difference between the classical case of metal-semiconductor contacts and the present case is the term that contains the polarization. Therefore, a new built-in potential denominated as the "apparent built-in potential" $V_{bi}'$ can be redefined in order to consider also the polarization charge contribution:

$$V_{bi}' = V_{bi} \pm \frac{P}{\varepsilon_0 \varepsilon_{st}} \delta \qquad (5)$$

As seen from (4) and (5) the built-in potential will apparently increase for positive polarization charge and will decrease for negative polarization charge, comparing with the built-in potential $V_{bi}$ in the normal metal–semiconductor case which is given by:[35,45]

$$V_{bi} = \Phi_B^0 - \frac{kT}{q} Ln\left(\frac{N_V}{p(T)}\right) \qquad (6)$$

where $\Phi_B^0$ is the potential barrier height at zero applied field; $k$ is the Boltzmann's constant; $T$ is the temperature, assumed constant; $N_V$ is the density of states in the valance band; $p(T)$ is the hole concentration at the temperature $T$. It has to be underlined that $p(T)$ is the concentration of free holes at the given temperature.[33]

The depletion width as well as all characteristic quantities of the Schottky contact can now be calculated in a similar way as in the case of a conventional Schottky contact, keeping in mind that the built-in potential must be replaced with the apparent built-in potential $V_{bi}'$:

$$w = \sqrt{\frac{2\varepsilon_0 \varepsilon_{st}(V + V_{bi}')}{qN_{eff}}} \qquad (7)$$

The maximum electric field at the interface will be:

$$E_m = \sqrt{\frac{2qN_{eff}(V + V_{bi}')}{\varepsilon_0 \varepsilon_{st}}} \pm \frac{P}{\varepsilon_0 \varepsilon_{st}} \qquad (8)$$

The plus sign applies when the polarization charge has the same sign with the fixed charge in the SCR. In the present case the charge in the SCR is negative, thus the polarization charge has the same sign in case of the reverse-biased contact (plus polarity on the electrode). The minus sign in (8) applies when the polarization charge is of opposite sign to the fixed charge in the SCR. This is the case of the forward-biased contact (minus polarity on the electrode). It has to be noted that (8) was deduced in the approximation that at any voltage the depletion width $w$ is significantly larger than the thickness of the interface layer $\delta$ ($w >> \delta$). $\delta$ is estimated to of the order of unit cell.[1] For standard values of $\varepsilon_{st} = 200$ and $N_{eff} = 10^{20}$ cm$^{-3}$ [1,6,16,32], the depletion width calculated with (7) is about 5 nm even for $V_{bi}' = 0.1$ V, and increases rapidly with the applied voltage reaching about 16 nm at 1 V. Therefor, the polarization term in (8) will bring a constant contribution to the maximum electric field at the interface, the polarization charge being already inside the SCR at zero bias.

Finally, the specific capacitance of the depleted region is:

$$C = \frac{\varepsilon_0 \varepsilon_{st}}{w} = \sqrt{\frac{q\varepsilon_0 \varepsilon_{st} N_{eff}}{2(V + V_{bi}')}} \tag{9}$$

The equations (7) and (9) for the depletion layer width and capacitance, respectively, are similar with those for semiconductor Schottky contacts except that the normal built-in potential $V_{bi}$ is replaced now by the apparent built-in potential $V_{bi}'$. In case of the maximum field (8) the difference is more important due to a constant term that polarization adds to the term proportional to $(V + V_{bi}')^{1/2}$.

The influence of the ferroelectric polarization on the built-in potential and consequently on all specific quantities of the Scottky contact is very important. The main consequence is that the apparent built-in potential given by (5) will have different values at the two interfaces of a MFM structure, even in the case of perfectly symmetric electrodes. That implies different band-bending at the two electrodes.

Another important discussion point is the electric field at the interface, which now becomes dependent on the value of ferroelectric polarization. According to (8) the maximum electric field at the interface can be increased or reduced depending on the sign of the polarization charge. The ferroelectric non-linearity is expected to have important effect on the interpretation of the I-V and C-V characteristics. These will be analyzed in detail in the following paragraphs.

## Current-voltage (I-V) characteristic

I-V characteristics are a convenient experimental way to investigate transport mechanisms in ferroelectric thin films. Referring to the specific case of MFM structures with Schottky contacts, the possible carrier injection mechanisms from the electrodes into the ferroelectric can be thermionic emission (TE), thermionic field emission (TFE), and field emission (FE). The prevalence of one or other of these mechanisms can be easily decided by comparing the characteristic energy $E_{00}$ with the $kT$ value, assuming that the temperature $T$ is fixed. The characteristic energy is given by:[33]

$$E_{00} = \frac{qh}{4\pi}\sqrt{\frac{N}{\varepsilon_0 \varepsilon_{st} m^*}} \tag{10}$$

where $h$ is Planck's constant, $N$ the doping concentration, and $m^*$ the effective mass. Considering $N=p(T)$, $E_{00}$ is about 1 meV at $T=300$K. As $kT$ is about 26 meV at room temperature, the condition $E_{00} \ll kT$ is fulfilled and the thermionic emission is the dominant carrier injection mechanism. In this case the current density is given by:[35]

$$J = \frac{qN_V v_R}{1+\frac{v_R}{v_D}} \exp\left(-\frac{q}{kT}\left(\Phi_B^0 - \sqrt{\frac{qE_m}{4\pi\varepsilon_0 \varepsilon_{op}}}\right)\right) \tag{11}$$

where $v_R$ is an effective recombination velocity at the collecting electrode, $v_D$ is the diffusion velocity, $E_m$ is the electric field at the maximum barrier height position, and $\varepsilon_{op}$ is the dynamic (high frequency) dielectric constant. The image-force effect, which lowers the potential barrier under the applied field, was considered in the above formula.

There are two limiting cases that are determined by the mobility of the free carriers in the ferroelectric layer. First, if the carrier recombination is negligible ($v_R \ll v_D$), the emission is pure thermionic and the current value is strictly controlled by the barrier properties. The current density is given in this case by:

$$J = A^* T^2 \exp\left(-\frac{q}{kT}\left(\Phi_B^0 - \sqrt{\frac{qE_m}{4\pi\varepsilon_0\varepsilon_{op}}}\right)\right) \quad (12)$$

where $A^*$ is Richardson's constant.
Second, in the case of strong recombination ($v_R \gg v_D$), the current density is:

$$J = qN_V v_D \exp\left(-\frac{q}{kT}\left(\Phi_B^0 - \sqrt{\frac{qE_m}{4\pi\varepsilon_0\varepsilon_{op}}}\right)\right) \quad (13)$$

The diffusion velocity $v_D$ is proportional with the carrier mobility and for low mobility materials it can be smaller than the recombination velocity $v_R$. In other words, the bulk of the material cannot supply enough carriers at the collecting electrode and, as a result, the conduction is bulk limited even if the injection mechanism is thermionic emission over the barrier.

It was shown that the transition from one mechanism to another takes place when $v_R = v_D$.[35] A critical filed can be roughly defined from the above condition, as: [35]

$$E_m^{cr} = \frac{A^* T^2}{qN_V \mu} \quad (14)$$

When the maximum field at the interface is larger than $E_m^{cr}$ then the emission is purely thermionic and eq. (12) applies. For fields lower than $E_m^{cr}$ the diffusion theory apply and eq.(13) should be used.

Unfortunately, the value of this field can not be accurately estimated in the particular case of ferroelectrics due to the lack of information regarding the value of the Richardson constant. The values of the carrier mobility in ferroelectrics are also not very reliable since the reported values are spread over more than two orders of magnitude in the 1–400 cm$^2$/Vs range.[46-50] Considering a mobility of 100 cm$^2$/Vs for the PZT materials, and Richarson's constant for the free electron of 120 A/cm$^2$K$^2$, the critical field can be estimated to be about 30 kV/cm. The effective density of states in the valance band $N_V$ was computed using the mass of the free electron. Referring to the maximum field at the interface, given by eq. (8), it can be observed that the polarization term contribute with a field of about 500 kV/cm for a polarization value of 10 μC/cm$^2$ and $\varepsilon_{st} = 200$. Further on it will be assumed that the maximum field at the metal-ferroelectric interface and at zero applied voltage calculated using (8) is higher than the critical field given by (14). Therefore, it will be assumed that the emission is purely thermionic and that (12) applies.

In the framework of the Schottky emission for a semiconductor rectifying contact the electric field in (11)–(13) must be considered the field given by (8) without the polarization term. Substituting (8) into (12) it can be easily deduced that ln($J$) is proportional to $(V+V_{bi})^{1/4}$:

$$\ln(J) = b(V + V_{bi}')^{1/4} \quad (15)$$

The slope $b$ of this representation is given by:

$$b = \frac{q}{kT}\sqrt[4]{\frac{q^3 N_{eff}}{8\pi^2 \varepsilon_0^3 \varepsilon_{op}^2 \varepsilon_{st}}} \quad (16)$$

The above equation is usually used to determine N$_{eff}$ for semiconductor Scottky contacts as well as for ferroelectrics. [35,36,51] We will try in the following to discuss the conditions under which the (15) and (16) can be applied and which are the limitations.

First, the polarization enters as a constant term in (8) and results in a deviation from the linear behavior of (15). It can be shown that, with the typical values mentioned above, the constant term in (8) is about 500 kV/cm while the voltage term reaches 1.4 MV/cm at 1 V. Therefore, it can be easily assumed that for usual applied voltage ranges Eq. (15) is valid. The physical meaning is that the constant polarization term in (8) becomes negligible when the depletion region extends well within the bulk of the film and the contribution of the fixed charges from the SCR (filled traps or ionized impurities) to the maximum field at the interface becomes dominant compared with the contribution of the bound polarization charges.

Second, both static and dynamic dielectric constants occur in (16). Compared with the classical case of semiconductors, when the difference between the static and the dynamic dielectric constant is not too large, in the case of ferroelectrics the dynamic dielectric constants is very low compared to the static value. For instance, the dynamic dielectric constant determined from optical transmission-reflectance measurements is about 6.5 and is orders of magnitude lower than the static one determined from direct impedance measurements.

The equations (16) can also be applied to (13) but using the following representation:

$$\ln\left(\frac{J}{\sqrt{V+V_{bi}'}}\right) = b(V+V_{bi}')^{1/4} \qquad (17)$$

This relation was derived using $v_D = \mu E_m$, with $E_m$ given by (8) and neglecting the constant term containing the ferroelectric polarization. A similar representation is obtained in the framework of Schottky's diffusion theory for metal-semiconductor rectifying contacts.[35] If the exponential factor in (13) is dominant, then it increases very fast with the voltage, and the differences between (15) and (17) will be not so large. The slope will be approximately the same. This is valid if $N_{eff}$ is large. Otherwise, a difference between the two slopes can exist, which is determined by the voltage dependence of the pre-exponential factor in (13). If both (15) and (17) are giving straight lines it will be difficult to decide which $N_{eff}$ value is the correct one as long as the exact value of the critical field (14) is not known for the investigated sample. The knowledge of the actual value of carrier mobility becomes a critical issue in this case.

Regarding (13) it can be seen that in the case of "semi-insulating" films and if the barrier lowering is negligible the I-V characteristic can have an apparent ohmic behavior if $v_D = \mu E$ and $E = V/d$, with $d$ being the film thickness. This ohmic behavior is possible in case of negligible band-bending in the ferroelectric.

In the case of a MFM structure, modeled as two back-to-back Schottky contacts, the current will be always limited by the reverse-biased contact. That can explain the often observed asymmetries between the positive and negative branches of the I-V characteristics. These asymmetries might originate in non-equivalent interfaces, occurring even when both top and bottom electrodes are made of the same material. It is highly probable that due to different processing conditions the potential barrier heights and/or the thickness of the interface layers are different. A non-uniform distribution of $N_{eff}$ within the film thickness might also be a possible explanation.

It is important to notice that the measurement of the true leakage current requires a pre-poling process in order to avoid currents due to the polarization reversal. This is usually performed by applying a suitable constant field, higher than the coercive field, and of the same polarity as the voltage intended to be used later on for the I-V measurement. In this way the polarization value and direction are set and no reversal currents occur during the measurement, assuming that the hysteresis curve is ideal as presented in Fig. 2.

## *Capacitance-voltage (C-V) characteristic*

The influence of the ferroelectric properties on the relevant quantities of the Schottky contact that can be extracted from C-V measurements is discussed below.

### The built-in potential

One of the most critical problems of any Schottky model, including the ferroelectric case, is the estimation of the built-in potential $V_{bi}$. This estimation is based on (9) and it is experimentally performed by measuring the C-V characteristic. $V_{bi}$ is directly obtained from the intercept of the plot $1/C^2$ versus applied voltage $V$. Unfortunately, this is valid only in the case of uniform doping and in the absence of reversible polarization. In the ferroelectric case it is not correct to estimate $V_{bi}$ from the intercept of the $1/C^2$ vs. $V$. This is mostly due to the way the C-V measurements are performed, i.e. by sweeping the applied bias from minimum (usually $-V$) to maximum ($+V$) and back again to $-V$. The polarization reversal occurs during the voltage sweep, therefor the conditions of the Schottky model are altered at voltages near the coercive values. This is of serious concern for ferroelectrics with hysteresis loops that are far from ideality. In this case the polarization reversal takes place on a wide voltage range, including zero. Therefore, (9) cannot be rigorously applied to determine the built-in voltage as in the case of normal semiconductors. Only in the case of an ideal hysteresis curve it might be possible to make a crude estimation for the apparent built-in potential $V_{bi}'$, and only in the absence of other built-in (internal) biases that can add to the one induced by the band-bending near the interface. Such internal potentials can occur due to the non-uniform distribution of space charges inside the structure. In most cases we deal with non-ideal polycrystalline ferroelectrics, therefor the $V_{bi}'$ estimated from (9) as for uniform doping is far from being correct and can not be directly used in (15).

### Charge density

In the frame of Schottky theory by taking the derivative of (8) the doping profile can be extracted according to the following relation: [35]

$$N_{dop} = \frac{2}{q\varepsilon_0\varepsilon_{st}\left[d(1/C^2)/dV\right]} \quad (18)$$

The depletion width can be calculated from the measured capacitance using equation (9), thus a pair $N_{dop} - w$ can be obtained for each C-V point. The graphic representation is the doping profile. If $N_{dop}$ is constant with $w$ then the doping is uniform within the film.

However, it is important to note that $N_{dop}$ is the density of the free carriers at the given temperature.[33] As it was emphasized by Schröder in ref. 33, due to the particularities of the direct capacitance measurements, only the free carriers can follow the small ac voltage over-imposed to the dc bias, and not the fixed charge within the depleted region. $N_{dop}$ is equivalent to the space charge density in the depleted region only if the shallow dopants are completely ionized at room temperature (RT) and the trap concentration is negligible. The existence of traps in a high concentration can cause difficulties in estimation of $N_{dop}$ because they might not follow the ac voltage at high frequency.

It was shown that the capacitance of a Schottky diode with an important trap density is:[52]

$$C = \left(\frac{\varepsilon_0\varepsilon_{st}N_{dop}}{2(V+V_{bi})}\right)^{1/2}\left(1 + \frac{N_T}{p(T)}\frac{e_n^2}{e_n^2 + \omega^2}\right) \quad (19)$$

where $N_T$ is the trap concentration, $e_n$ is the trap emission rate (the trap is assumed for electrons), and $\omega = 2\pi f$, with $f$ being the frequency of the ac probing voltage used in the C-V measurement. If $e_n$ is very small at the given temperature, the trap contribution to the measured capacitance becomes negligible. The second parenthesis in (19) can be approximated by unity and the capacitance is the same as in the simple case of no traps. We underline that this behavior is possible in the case of deep levels in wide-gap semiconductors, when the thermal emission rate $e_n$ from the trap level is very small compared with the frequency used in C-V measurements. In this case the charge concentration determined from the C-V characteristics is not the same as the space charge density of the depleted region determined from the I-V measurements. The charge concentration extracted from the C-V characteristic is the concentration of free carriers at the given temperature, and not the effective fixe charge $N_{eff}$ in SCR. Therefore, from the C-V measurements $p(T)$ occurring in (6) is in fact calculated, whereas the major role in transport mechanism is plaid by $N_{eff}$, occurring in (8) and (16). Parenthetically it has to be noted that the free carrier concentration is controlled by the deep level if it pins the Fermi level.[39] The hole concentration $p(T)$ evaluated from C-V measurements should have a strong temperature dependence in this case. If the deep level is just an electron trap, with no influence on the hole concentration, then the value *p(T)* evaluated from C-V characteristic should be temperature independent, providing that the shallow acceptors are fully ionized at RT.

As ferroelectrics are far from being free of structural defects, the analysis should be rather performed using the complex system with traps than the idealized case of simple Schottky contacts.

### Depletion layer width

The second important quantity of the Schottky contact is the depletion layer width calculated using (7). In usual cases the trap concentration is much smaller than the acceptor concentration ($N_T << N_A$). In the present case it was assumed that the trap concentration is higher than the acceptor concentration. This fact can bring some complications related with the band bending, which is the same for all levels in SCR, but the point where the deep trap level crosses the Fermi level may be situated at a shorter distance from the interface than the width of the SCR. This fact depends on its energetic position compared with the Fermi level. This situation is sketched in Fig. 4. Compared with Fig. 3, where the Fermi energy is assumed to be pinned by the deep level, in Fig. 4 the deep level is far from the Fermi energy. In this case the charge density is no longer uniformly distributed in the depletion region. Between the metal-ferroelectric interface and the above intersection point the space charge will be dominated by the trap level, which is now filled because it is below the Fermi level and contributes with a significant negative charge. Between the intersection point and the edge of the space charge region the space charge is dominated by the ionized acceptors. In this region the trap level is empty, being located above the Fermi level, and its contribution to the space charge is zero. The integration of (3) will give in this case:

$$V + V_{bi} = \frac{q}{2\varepsilon_0 \varepsilon_{st}} \left[ N_T w_1^2(V) + N_A \left( w_2^2(V) - w_1^2(V) \right) \mp 2\frac{P}{q} \delta \right] \quad (20)$$

$w_1(V)$ defines the point where the trapping level crosses the Fermi level, while $w_2(V)$ defines the normal depletion width. Both quantities are voltage dependent thus, is not possible to extract the voltage dependence for both of them. An effective depletion width can be defined by:

$$w_{eff}(V) = \sqrt{\frac{w_1^2(N_T - N_A) + w_2^2 N_A}{N_{eff}}} \quad (21)$$

Obviously, the effective charge density is the sum of the acceptor and trap concentrations $N_{eff} = N_A + N_T$. If $N_T >> N_A$ then $N_{eff} \sim N_T$ and $w_{eff}$ is approximately equal with $w_1$. The depletion width is controlled by the deep trap level in this case. If $N_T << N_A$ then $w_{eff} \sim w_2$ and the depletion width is controlled by the shallow acceptor level.

It is also important to understand the role played by this space charge non-uniformity. From the point of view of the electric field at the interface, which plays the major role in the charge injection mechanism, the dominant role is played by the trap density $N_T$ near the interface, thus (8) must be evaluated using $N_{eff} \sim N_T$. From the point of view of the measured capacitance, the dominant role is played by the free carriers as discussed in the previous paragraph. If all the sallow impurities are ionized at the given temperature then $p(T) = N_A$ and $N_{eff}$ should be replaced by $N_A$ in (9).[43,44] We underline that both the shallow acceptors and the deep levels are considered uniformly distributed throughout the film thickness. The non-uniform charge density in the depleted region is only apparent and is dependent on the balance between the concentrations $N_A$ and $N_{eff}$ and on the energetic depth of the deep level.

## *Conclusions*

A model for metal-ferroelectric-metal structures was developed. It assumes that the ferroelectrics are wide gap p-type semiconductors and adapts the already known theories for metal-semiconductor rectifying contacts to the metal-ferroelectric contact by considering: i) the ferroelectric polarization as a surface charge near the electrode interface; ii) an ideal hysteresis loop; iii) a thin layer with zero polarization at the metal-ferroelectric interface; iv) a deep trap level of high concentration; and v) both static and dynamic dielectric constants.

The model shows the influence of the polarization on the built-in potential and on the maximum field at the interface. The consequences of the proposed model on the interpretation of the current-voltage (I-V) and capacitance-voltage (C-V) characteristics were discussed in detail, along with the limitations of model in case of thermionic emission. It was shown that apparent space charge non-uniformity can occur in the depleted region if the trap level is deep in the band gap and its concentration is higher than the acceptor concentration. The consequences of this fact on evaluation of the electric field and MFM capacitance were also discussed. The possibility to obtain different values of the charge density from the analysis of the I-V and C-V characteristics was underlined. The present model can be the basis for a more general description of the electronic properties of MFM structures, including charge transport through very thin ferroelectric films.[53]

## *Acknowledgements*

This work was partly supported by Volkswagen Stiftung in the project 'Nano-sized Ferroelectric Hybrids' under contract No. I/77737 and I/80897 and partly by NATO under project SfP-971970 and grant CP(RO)04/C/2001/PO.
The authors are grateful to Prof. Ulrich Goesele for the useful discussion.

## *References*

1. J. F. Scott, Ferroelectric Memories, in Advanced Microelectronics Series, edited by K. Itoh and T. Sakurai (Springer-Verlag Berlin Heidelberg, 2000).
2. K. Uchino, Ferroelectric Devices, (Marcel Dekker, New York, 2000)
3. M. Grossman, O. Lohse, D. Bolten, U. Boettger, and R. Waser, J. Appl. Phys. **92**, 2688, (2002); J. Appl. Phys. **92**, 2680 (2002).
4. T. Mihara and H. Watanabe, Jpn. J. Appl. Phys. **34**, 5664 (1995); Jpn. J. Appl. Phys. **34**, 5674 (1995).
5. P. W. Boom, R. M. Wolf, J. F. M. Cillessen, and M. P. C. M. Krijn, Phys. Rev. Lett. **73**, 2107 (1994)
6. J. C. Shin, J. Park, C. S. Hwang, and H. J. Kim, J. Appl. Phys. **86**, 506, (1999)
7. C. Shudhama, A. C. Campbell, P. D. Maniar, R. E. Jones, R. Moazzami, C. J. Mogab, and J. C. Lee, J. Appl. Phys. **75**, 1014 (1994)
8. I. Stolichnov and A. Tagantsev, J. Appl. Phys. **84**, 3216 (1998)
9. B. Nagarajan, S. Aggaarwal, T. K. Song, T. Sawhney, and R. Ramesh, Phys. Rev. B **59**, 16022 (1999)
10. C. Hwang, B. T. Lee, C. S. Kang, K. H. Lee, H. Cho, H. Hideki, W. D. Kim, S. I. Lee, and M. Y. Lee, J. Appl. Phys. **85**, 287 (1999)
11. Y. S. Yang, S. J. Lee, S. H. Kim, B. G. Chae, and M. S. Jang, J. Appl. Phys. **84**, 5005 (1998)
12. B. Nagarajan, S. Aggarwal, and R. Ramesh, J. Appl. Phys. **90**, 375 (2001)
13. S. G. Yoon, A. I. Kingon, and S. H. Kim, J. Appl. Phys. **88**, 6690 (2000)
14. S. Bhattacharyya, A. Laha, and S. B. Krupanidhi, J. Appl. Phys. **91**, 4543 (2002)
15. H. Schroeder, S. Schmitz, and P. Meuffels, Appl. Phys. Lett. **82**, 781 (2003)
16. J. D. Baniecki, T. Shioga, K. Kurihara, and N. Kamehara, J. Appl. Phys. **94**, 6741 (2003)
17. S. Gopalan, V. Balu, J. H. Lee, J. H. Han, and J. C. Lee, Appl. Phys. Lett. **77**, 1526 (2000)
18. S. T. Chang and J. Y. Lee, Appl. Phys. Lett. **80**, 655 (2002)
19. S. Sadashivan, S. Aggarwal, T. K. Song, R. Ramesh, J. T. Evans Jr., B. A. Tuttle, W. L. Warren, and D. Dimos, J. Appl. Phys. **83**, 2165 (1998)
20. A. Q. Jiang, J. F. Scott, M. Dawber, and C. Wang, J. Appl. Phys. **92**, 7656 (2002)
21. M. V. Raymond and D. M. Smyth, J. Phys. Chem. Solids **57**, 1507 (1996)
22. S. Aggarwal and R. Ramesh, Ann. Rev. Materials, **28**, 463 (1198)

23. V. K. Yarmarkin, B. M. Gol'tsman, M. M. Kazanin, and V. V. Lemanov, Phys. Of the Solid State, **42**, 511 (2000)
24. J. F. Scott, C. A. Araujo, B. M. Melnick, L. D. McMillan, and R. Zuleeg, J. Appl. Phys. **70**, 382 (1991)
25. M. Dawber, and J. F. Scott, Appl. Phys. Lett. **76**, 1060 (2000)
26. C. H. Lin, P. A. Friddle, C. H. Ma, A. Daga, and H. Chen, J. Appl. Phys. **90**, 1509 (2001)
27. B. M. Gol'tsman, V. K. Yarmarkin, and V. V. Lemanov, Phys. Of the Solid State, **42**, 1083 (2000)
28. F. M. Pontes, D. S. L. Pontes, E. R. Leite, E. Longo, A. J. Chiquito, P. S. Pizani, and J. A. Varela, J. Appl. Phys. **94**, 7256 (2003)
29. L. Lee, C. H. Choi, B. H. Park, T. W. Noh, and J. K. Lee, Appl. Phys. Lett. **72**, 3380 (1998)
30. D. P. Chu, Z. G. Zhang, P. Migliorato, B. M. McGregor, K. Ohashi, K. Hasegawa, and T. Shimoda, Appl. Phys. Lett. **81**, 5204 (2002)
31. H. Z. Jin and J. Zhu, J. Appl. Phys. 92, J. Appl. Phys. **92**, 4594 (2002)
32. J. McAneney, L. J. Sinnamon, R. M. Bowman, and J. M. Gregg, J. Appl. Phys. **94**, 4566 (2003)
33. D. K. Schroeder, Semiconductor material and device characterization, (Wiley-Interscience, New York, 1998)
34. P. Braeunlich, Thermally stimulated relaxation in solids, (Springer, Berlin, 1979)
35. S. M. Sze, Physics of Semiconductor Devices, 2$^{nd}$ ed. (John Wiley & Sons, 1981), Chaps 7 and 10.
36. K. C. Kao and W. Hwang, Electronic Transport in Solids, International series in the Science of the Solid State, vol. 14. General editor B.R. Pamplin (Pergamon Press, Oxford, 1981)
37. R. Kretschmer and K. Binder, Phys. Rev. B **20**, 1065, (1979)
38. M. E. Lines and A. M. Glass, Principles and Applications of Ferroelectrics and Related Materials (Clarendon Press, Oxford, 1977)
39. H. Shen, F. C. Rong, R. Lux, J. Pamulapati, M. Taysing-Lara, M. Dutta, and E. H. Poindexter, Appl. Phys. Lett. **61**, 1585 (1992)
40. Y. H. Chen, Z. Yang, Z. G. Wang, and R. G. Li, Appl. Phys. Lett. **72**, 1866 (1998)
41. Y. H. Chen, Z. G. Wang, and Z. Yang, J. Appl. Phys. **87**, 2923 (2000)
42. X. M. Lu, F. Schlaphof, S. Grafstroem, C. Loppacher, and L. M. Eng, Appl. Phys. Lett. **81**, 3215 (2002)
43. D. Goren, N. Amir, and Y. Nemirovsky, J. Appl. Phys. **71**, 318 (1992)
44. D. Goren and Y. Nemirovsky, J. Appl. Phys. **77**, 244 (1995)
45. K. Shiojima and J. M. Woodall, J. Vac. Sci. Technol. B **17**, 2030 (1999)
46. S. Zafar, R. E. Jones, B. Jiang, B. White, V. Kaushik, and S. Gillespie. Appl. Phys. Lett. **73**, 3533 (1998)
47. H. H. Wang, F. Chen, S. Y. Dai, T. Zhao, H. B. Lu, D. F. Cui, Y. L. Zhou, Z. H. Chen, and G. Z. Yang, Appl. Phys. Lett. **78**, 1676 (2001)
48. D. Nagano, H. Funakubo, K. Shinozaki, and N. Mizutani, Appl. Phys. Lett. **72**, 2017 (1998)
49. Y. S. Yang, S. J. Lee, S. H. Kim, B. G. Chae, and M. S. Jiang, Jpn. J. Appl. Phys. **36**, 749 (1997)
50. D. C. Lupascu, Phys. Rev. B **70**, 184124 (2004)
51. I. Boerasu, L. Pintilie, M. Pereira, M. I. Vasilevskyi, and M. J. M. Gomes, J. Appl. Phys. **93**, 4776 (2003)
52. G. Vincent, D. Bois, and P. Pinard, J. Appl. Phys. **46**, 5173 (1975)
53. J. Rodriguez Contreras, H. Kohlstedt, U. Poppe, R. Waser, C. Buchal, and N. A. Pertsev, Appl. Phys. Lett. **83**, 4595 (2003).

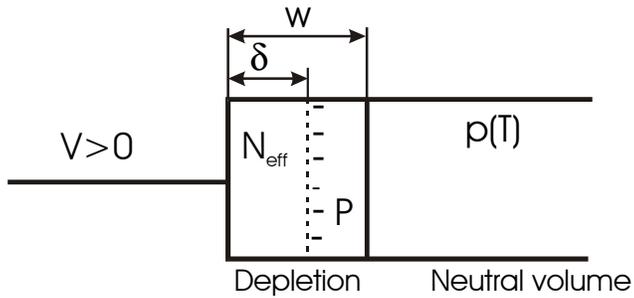

Figure 1 Schematic model of the metal-ferroelectric interface. Ferroelectric polarization is considered as a infinite thin surface charge layer of *P* charge density, $\delta$ is the thickness of the interface layer, *w* is the depletion layer width, $N_{eff}$ the effective charge density in the depleted layer, and *p(T)* is the hole concentration in the neutral volume. In a MFM structure there will be two sheets of polarization charge, of opposite sign, located near the two electrode-ferroelectric interfaces.

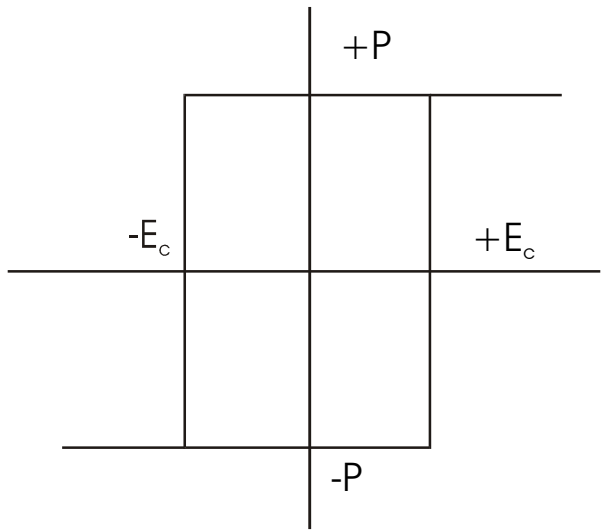

Figure 2 Hysteresis loop of an ideal ferroelectric material. Spontaneous polarization *P* is equal to the remanent polarization and switching is performed only at coercive field $\pm E_c$.

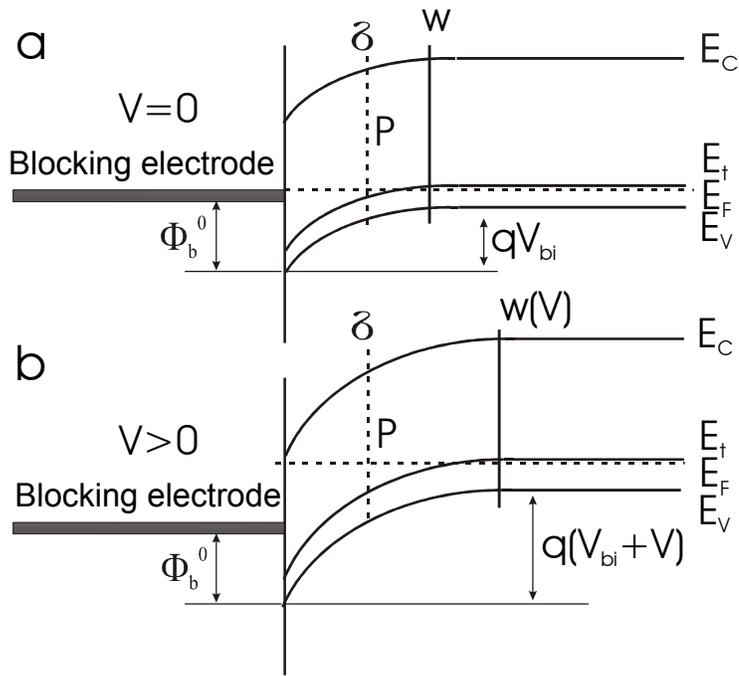

Figure 3 Schematic band diagram of the metal-ferroelectric reverse-biased contact at (**a**) V=0 and (**b**) at V>0.

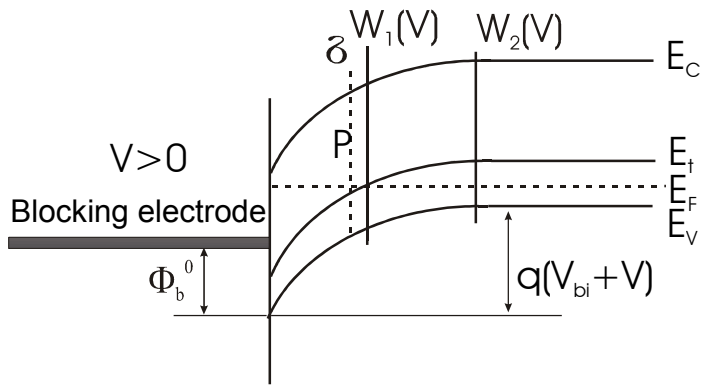

Figure 4 Schematic band diagram of the reverse-biased metal-ferroelectric contact showing the occurrence of the apparent non-uniformity in the space charge distribution.